\documentclass[conference,review,authorversion]{IEEEtran} 
\IEEEoverridecommandlockouts

\AtBeginDocument{%
  }
\usepackage{amsmath,amsfonts}
\usepackage{algorithmic}
\usepackage{graphicx}
\usepackage{textcomp}
\usepackage{xcolor}
\usepackage{caption}
\usepackage{subcaption}
\usepackage{booktabs}
\usepackage{graphics}\usepackage{listings}
\usepackage{color}
\usepackage{tcolorbox}
\usepackage{multirow}
\usepackage[ruled, vlined]{algorithm2e}
\usepackage{verbatim}
\usepackage{mdframed}
\usepackage{tabularx}
\usepackage{arydshln} % for dashline
\usepackage{tablefootnote}
\usepackage{tikz}
\usetikzlibrary{tikzmark}
\usepackage{threeparttable}
\usepackage{pifont}
\usepackage{adjustbox}
\usepackage{url}

\usepackage{hyperref}
\hypersetup{
    colorlinks=false,
}
\usepackage{balance}

\definecolor{dkgreen}{rgb}{0,0.6,0}
\definecolor{gray}{rgb}{0.5,0.5,0.5}
\definecolor{mauve}{rgb}{0.58,0,0.82}

\newcommand*\circled[1]{\tikz[baseline=(char.base)]{
    \node[shape=circle, draw, inner sep=1pt, minimum size=8pt, font=\scriptsize] (char) {#1};}}
\newcommand{\picircle}[1]{\textcircled{\scriptsize #1}}
            
\definecolor{highlightcolor}{RGB}{255, 255, 0} % Yellow color

\newcommand{\phead}[1]{\vspace{1mm} \noindent {\bf #1}}

\newcommand{\rqbox}[1]{\begin{tcolorbox}[left=2pt,right=2pt,top=2pt,bottom=2pt,colback=gray!5,colframe=gray!40!black,before skip=2pt,after skip=2pt]#1\end{tcolorbox}}

\definecolor{brinkpink}{rgb}{0.98, 0.38, 0.5}

\newcommand{\anran}[1]{\textcolor{orange}{{[An Ran: #1]}}}

\newcommand{\yiwen}[1]{\textcolor{violet}{{[Yi Wen: #1]}}}

\newcommand{\djk}[1]{\textcolor{purple}{{[DJ: #1]}}}

\begin{document}

%\title{Benchmarking Open-Source Large Language Models For Log Level Suggestion}
\title{Studying and Benchmarking Large Language Models For Log Level Suggestion}

% --- IEEE Template
%\author{\IEEEauthorblockN{Yi Wen Heng, Tse-Hsun (Peter) Chen}
%\IEEEauthorblockA{\textit{Department of Computer Science and Software Engineering}
%{\textit{Concordia University}} \\
%Montreal, Canada \\
%\{he\_yiwen, peterc\}@encs.concordia.ca}}
%\maketitle
% --- IEEE Template

\author{%%%% author names
    \IEEEauthorblockN{Yi Wen Heng\textsuperscript{1}, Zeyang Ma\textsuperscript{1}, Zhenhao Li\textsuperscript{2}, Dong Jae Kim \textsuperscript{1}, Tse-Hsun (Peter) Chen\textsuperscript{1}}% first author
    % duplicate the line above as many times as needed to list all authors
    \IEEEauthorblockA{\textit{\textsuperscript{1}Software PErformance, Analysis, and Reliability (SPEAR) lab, Concordia University, Montreal, Canada}}
    \IEEEauthorblockA{\textit{\textsuperscript{2}York University}}

    \IEEEauthorblockA{he\_yiwen@encs.concordia.ca, m\_zeyang@encs.concordia.ca, zhenhao.li@ieee.org}
    \IEEEauthorblockA{k\_dongja@encs.concordia.ca, peterc@encs.concordia.ca}
}

% --- ACM Template
% \keywords{language models, prompting, few-shot learning, prompt-tuning, software engineering}
\maketitle
% --- ACM Template

% --- IEEE Template
% \begin{IEEEkeywords}
% language models, prompting, few-shot learning, fine-tuning, log level, empirical study
% \end{IEEEkeywords}
% --- IEEE Template

\begin{abstract}
   Large Language Models (LLMs) have become a focal point of research across various domains, including software engineering, where their capabilities are increasingly leveraged. Recent studies have explored the integration of LLMs into software development tools and frameworks, revealing their potential to enhance performance in text and code-related tasks. Log level is a key part of a logging statement that allows software developers control the information recorded during system runtime. Given that log messages often mix natural language with code-like variables, LLMs' language translation abilities could be applied to determine the suitable verbosity level for logging statements. In this paper, we undertake a detailed empirical analysis to investigate the impact of characteristics and learning paradigms on the performance of 12 open-source LLMs in log level suggestion. We opted for open-source models because they enable us to utilize in-house code while effectively protecting sensitive information and maintaining data security. We examine several prompting strategies, including Zero-shot, Few-shot, and fine-tuning techniques, across different LLMs to identify the most effective combinations for accurate log level suggestions. Our research is supported by experiments conducted on 9 large-scale Java systems. The results indicate that although smaller LLMs can perform effectively with appropriate instruction and suitable techniques, there is still considerable potential for improvement in their ability to suggest log levels.

\end{abstract}

% ------------Main sections------------

\section{Introduction}
\label{sec:introduction}

% \zhenhao{IEEE template if intent to submit to ICST 2025.}

%Logs are vital for software development and maintenance, providing critical runtime information that supports testing~\cite{Chen2018AnAA, Chen2017AnalyticsDrivenLT, Li2015GatedGS}, debugging~\cite{Fu2014WhereDD, Yuan2010SherLogED, Zhao2017Log20FA}, failure diagnosis~\cite{Yuan2011, Zhou2019LatentEP, Schipper2019TracingBL, Su2022WhyMA, Zhang2021OnionII, Chen2022PathideaII}, program comprehension~\cite{Messaoudi2021LogbasedSF, Nagaraj2012StructuredCA, li2023they, Gadler2017MiningLT, Li2020TowardsPA}, and anomaly detection~\cite{Shin2021ATF, Yang2021PLELogSL, Zhang2019RobustLA, Zhao2021AnEI}. However, the massive volume of logs generated by modern software systems, often reaching tens of gigabytes or even terabytes daily~\cite{zhu2019, logram, Kernel}, necessitates labor-intensive manual processes for management and analysis.

%\peter{we should add some keywords like testing and quality assurance throughout the intro}

% \todo{ICST 2025 abstract due, Sept. 25.}
% \zhenhao{For the bib reference, (1) remove all the information except author, title, venue \& year, page (e.g., DOI, URL). The reference should be no more than 2 pages; (2) Use IEEETran (not plainnat and natbib, othersie the fontsize is too large)}.
%\todo{You have until Wednesday Oct 2, 2024, 11:59:59 PM AoE (Oct 3 7:59:59 AM your time) to submit papers.}

% ~\djk{You provide an example, but do not discuss anything meaningful after. }
% ~\yw{I changed the sequence to show the importance of logs}
Logs are invaluable for tracking system runtime behavior. 
Logs contribute to software quality assurance by monitoring system performance and reliability~\cite{Chen2017AnalyticsDrivenLT, Li2015GatedGS}, debugging~\cite{Fu2014WhereDD, Zhao2017Log20FA}, failure diagnosis~\cite{Yuan2011, Zhou2019LatentEP, Chen2022PathideaII}, program comprehension~\cite{Messaoudi2021LogbasedSF, li2023they, Gadler2017MiningLT, Li2020TowardsPA}, and anomaly detection~\cite{Shin2021ATF, Zhang2019RobustLA}. 
As an example, the logging statement {\tt LOG.debug("Task FINISHED, but concurrently went to state " + state);} is at the ``\textit{debug}'' level and contains the log message ``\textit{Task FINISHED, but concurrently went to state}'', and records the value for the {\tt state} variable. Such logging statements capture important runtime information for later analysis. 
%The following shows an example of logging statement, where ``debug", ``Task FINISHED, but concurrently went to state ", and ``state" represents the log level, log message, and log variable respectively.
%\fbox{\footnotesize LOG.debug("Task FINISHED, but concurrently went to state " + state);}

Despite their importance, the massive volume of logs generated by modern software systems, often reaching tens of gigabytes or even terabytes daily~\cite{zhu2019, Kernel, li2021studying}, poses significant challenges in log management and analysis, potentially bottlenecking quality assurance processes. 
To facilitate log management, verbosity levels (e.g., \texttt{trace}, \texttt{debug}, \texttt{info}, \texttt{warn}, and \texttt{error}) are employed to indicate the urgency and prioritization of log analysis. Logs can also be wrapped in log guards, such as \texttt{isDebugEnabled} or \texttt{isErrorEnabled}, which optimize performance by verifying whether specific logging levels are enabled before generating log messages. 

However, selecting appropriate verbosity levels in logging statements can be challenging due to a limited understanding of system runtime behaviors~\cite{Li2021, Oliner2012}. This often leads to incorrect log-level assignments, influenced by subjective interpretations or human error~\cite{Li2021}. Such misclassifications can result in critical messages being overlooked or trivial events being misrepresented, complicating log management and analysis efforts~\cite{Yuan2010SherLogED, Li2021} and imposing additional overhead on these processes~\cite{Li2021, Yuan2012CharacterizingLP}.
Tools like GitHub Copilot, which are powered by large language models (LLMs), offer a potential solution to mitigate these challenges by providing automated code improvement and competition, including log level suggestions~\cite{githubResponsibleGitHub}. Yet, despite the usefulness of these tools, companies such as Apple, Samsung, and Amazon have banned the use of AI tools powered by Large Language Models (LLMs) due to concerns about compromising proprietary code or sensitive information~\cite{forbesAppleJoins, forbesSamsungBans}.

In this paper, we conducted a comprehensive empirical evaluation of various open-source LLMs. Our focus on open-source models stems from the desire to enable users to leverage LLMs without concerns about data privacy. Given that log levels are essential for code improvement, we use log level suggestions as a key task to assess the performance of these models. %as one of the first evaluations. % to evaluate the performance of various types and sizes of open-source LLMs. 
We evaluated models of different sizes and architectures, including general-purpose language models like BERT and RoBERTa, as well as code-specific models such as CodeBERT and GraphCodeBERT.

For the experiment, we employed a benchmark dataset consisting of logging statements from nine large-scale, open-source Java systems. These systems cover a variety of domains and represent real-world software projects, making them well-suited for examining logging practices and log level predictions. We assessed each LLM using different learning paradigms, including zero-shot, few-shot, and fine-tuning approaches, to evaluate their effectiveness in suggesting the appropriate verbosity levels (e.g., debug, info, warn, error) for log statements.

We first preprocessed the dataset to extract logging statements alongside their corresponding source code, isolating the relevant code segments and log messages for each event. Then, we created prompts for the LLMs that included both the log message and the surrounding code context, aiming to simulate real-world logging scenarios faced by developers. Following prior studies~\cite{DeepLV,TeLL}, we use Accuracy, Area Under the Curve (AUC), and Average Ordinal Distance Score (AOD), to assess the performance of LLMs in suggesting log levels. We also explored how incorporating additional context, such as calling methods, influenced the model's accuracy in log level predictions. This experimental setup enabled us to benchmark both small and large models, providing insights into how model size, fine-tuning, and contextual information affect performance in practical software logging tasks. In summary, the contributions of our paper are as follows:

%We began by preprocessing the dataset to extract logging statements and their corresponding source code. Next, we created prompts for the LLMs that included both the log message and surrounding code context to simulate real-world scenarios. In line with prior research~\cite{DeepLV,TeLL}, we assessed LLM performance in log level suggestion using Accuracy, Area Under the Curve (AUC), and Average Ordinal Distance Score (AOD). We also examined how additional context, such as calling methods, impacted accuracy. This setup allowed us to benchmark both small and large models, offering insights into how model size, fine-tuning, and context affect performance in software logging tasks. In summary, the contributions of our paper are as follows\footnote{\peter{put it in reference}Our replication package is available at~\cite{replication_package}. \todo{Update the link} \zhenhao{Apart from GitHub, You can also try figshare (get the private share link, don't need to publish the uploaded data)}}:
\begin{itemize}
    % \item We discover several \peter{need to mention what are the trends}trends between Code-based and NLP-based LLMs, as well as between Fill-Mask and Text Generation LLMs, highlighting their differing effectiveness for log level suggestion tasks. 
    % \item \peter{please check this point. I don't really get what this means}
    % We discover that larger LLMs generally give better log level suggestions, but Fill-Mask models outperform them. Selecting a model trained on relevant data is more important than having more parameters. Fine-tuning improves results in 10 out of 12 models, and a fine-tuned Fill-Mask LLM performs nearly as well as state-of-the-art models.

    \item 
    We discover that Fill-Mask models such as GraphCodeBERT, when fine-tuned, can outperform larger models in log level suggestions. Fine-tuning using task-specific data leads to substantial improvements in accuracy and makes these models competitive with state-of-the-art methods.
    
    % \peter{check out and cite/reference this paper: Large Language Models Can Be Easily Distracted by Irrelevant Context}
    \item 
    We identify the impact of additional context, such as including the source code of calling methods, on LLM performance, noting that it can decrease accuracy and increase invalid outputs. The finding shows that adding undistilled information may negatively influence model output. %, as highlighted by the findings by Wang et al.~\cite{shi2023largelanguagemodelseasily}. 
    
    \item We highlight the critical role of task-specific data in optimizing LLMs and show that Text Generation LLMs are more effective when such data is unavailable. This lays the groundwork for future research focused on enhancing LLMs for code-related tasks.
\end{itemize}

\noindent\textbf{Paper organization.} 
%Write when sections are determined
Section~\ref{sec:background} introduces the background and related works to our study. Section~\ref{sec:datasrc} describes the overall design of our study. Section~\ref{sec:results} presents the study results. Section~\ref{sec:futureworks}  offers actionable insights and suggestions for future directions. Section~\ref{sec:threats} discusses the threats to validity. Section~\ref{sec:conclusions} concludes the paper.

\section{Background and Related Work}
\label{sec:background}

In this section, we discuss the background and related work of our study.

\subsection{Logging}
\phead{Log Level Suggestion.}
Prior works have investigated log-level suggestions using various machine learning and deep learning techniques~\cite{Li2016, DeepLV, MultiComponentLogLevel, TeLL}. Li et al.~\cite{Li2016} utilized the ordinal regression machine learning model to recommend appropriate log levels for logging statements. 
Ouatiti et al.~\cite{MultiComponentLogLevel} uses ordinal regression model and focuses on the performance of the log level suggestion that is trained separately multi-component software systems and their components. 
Li et al.~\cite{DeepLV} uses Ordinal Based Neural Networks to analyze syntactic context and message features of logging statements in order to suggest log levels.
Liu et al.~\cite{TeLL} applied graph neural networks to encode intra-block and inter-block features into code block representations, guiding log level suggestions.
% The studies below are less relevant, can comment them to save some space
%Moreover,~\citet{Tang2022} recognized that software evolution can influence log levels associated with surrounding features. They proposed an automated approach that mines Git histories and employs a Degree of Interest (DOI) model to adjust log levels based on the "interestingness" of surrounding code. 
%Additionally, Chen et al.~\cite{Chen2018AnAA} identified wrong verbosity as a logging code anti-pattern and developed a tool to detect and synchronize mismatches between static text and verbosity levels, enhancing it with an adapted version of the SZZ algorithm~\cite{boyuan2019}. Another prior study~\cite{Hassani2018} utilized Normalized Shannon's Entropy to determine logging statement levels based on the probability of phrase appearances in log messages. 

In contrast, our study uses open-source LLMs for log level suggestion to examine how different attributes and learning paradigms influence model performance. This approach makes use of readily available technology and offers insights into how various LLM characteristics and learning strategies can be effectively applied to log level suggestion.
% \peter{what is the difference between this and prior work?}

% \section{Related Work}
\label{sec:relatedwork}

\phead{Logging Statement Generation with LLMs.}
Some studies have explored logging statement generation using LLMs. For instance, Li et al.~\cite{GoStatic} proposed incorporating static context into code prompts using a self-refinement approach based on chain-of-thought (COT) prompts derived from static analysis. 
Xu et al.~\cite{UniLog} utilized Codex, a fine-tuned GPT language model~\cite{Chen2021EvaluatingLL}, to generate logging statements, examining both In-Context Learning and fine-tuning approaches. 
Unlike our approach, which utilizes open-source language models, Xu et al.~\cite{UniLog} employed several models from the GPT-3 series for comparison purposes. Another close work to our study is by Li et al.~\cite{Li2023ExploringTE}, where they conducted an empirical study on the ability of LLM in logging statement generation by creating a dataset of unseen code, selected 11 general-purpose, logging-specific, and code-based LLMs, and used prompt instruction on the LLMs to generate result. While their study covered the In-Context Learning paradigm, it did not include fine-tuning. 

% \peter{this needs to be revised. We should say our purpose is on benchmarking and studying how these factors affect the results (basically what the intro says), and what it means for future studies (inspire future code enhancement studies)}

% While previous work has primarily concentrated on automatically generating logging statements, our research distinguishes itself by focusing on suggesting appropriate log levels. Specifically, we leverage the capabilities of LLMs to analyze both the context of log messages and the semantics of the code within the method’s source code, thereby providing more contextually appropriate log level recommendations. Our findings not only provide insights into improving log level suggestions but also offer a foundation for future studies to explore code enhancement techniques via LLMs.

While previous research has largely focused on the automatic generation of logging statements and integrating LLM-based tools into software development practices, our study centers on benchmarking LLMs for log level suggestions. We aim to examine how factors such as model size, task types, and learning paradigms influence the outcomes. Our findings not only shed light on improving log level suggestions but also lay the groundwork for future studies on enhancing code quality through LLMs.
% ~\zeyang{I feel this paragraph is too similar to the last paragraph of Log Level Suggestion. We should highlight we focusing on log level here.}

% \peter{again, what is the difference between this and prior work?}

%Large Language Models have been used in studies to generate log statements\cite{UniLog, Li2023ExploringTE, Mastropaolo2022UsingDL}. However, in this study, we aim to determine which LLMs may be more suitable for suggesting log-levels. For instance,code-based LLMs like CodeBERT might be more suitable than NLP-based LLMs, given that our dataset comprises code extracted from systems. Additionally, we are investigating whether the training objective contributes to higher accuracy; for example, does CodeBERT for Java offer greater accuracy since it is specifically trained with Java code snippets? We are also comparing different types of LLMs, like Fill-Mask and Generative models, to see which is better at predicting log levels accurately. Similar to~\cite{UniLog} that compares which of the techniques, In-Context Learning and Fine-tuning, is more effective in adapting pre-trained models to new tasks, we also extend our study to find out how much training is needed for LLMs to accurately predict log levels.

\subsection{Large Language Models}
In recent years, advancements in natural language processing (NLP) have been transforming various fields, including software engineering, highlighting the importance of LLMs in code-related work, emphasizing the significance of LLMs in coding tasks, which are essential for enhancing artificial intelligence in Software Engineering (AI4SE). Recent research has extensively explored the integration of LLMs into software engineering tools and processes to enhance development practices and advance both academic and industry applications. Studies have examined how LLMs improve tasks such as logging statement generation~\cite{GoStatic, UniLog, Chen2021EvaluatingLL} and log parsing~\cite{Ma_2024, LILAC}.
Despite these advancements, several areas remain unclear and warrant further investigation:

\subsubsection{\textbf{Characteristics of LLMs}}
\label{LLMCharacteristics}
The effectiveness of various LLMs in performing specific tasks remains an area of active research. The following characteristics of LLMs contribute to their performance and influence how well they can handle different applications:

\phead{Task Types.} 
Text Generation models, like those trained with autoregressive techniques (e.g., LLaMA 2), excel in generating coherent and contextually relevant text by modeling long-range dependencies and ensuring sequential consistency from given prompts~\cite{LLaMA}. 
In contrast, Fill-Mask models like BERT and RoBERTa are trained by masking random tokens in a sentence and predicting them based on both left and right context. This bidirectional approach helps the model leverage context from both directions, making them particularly effective for tasks requiring deep contextual understanding~\cite{BERT, RoBERTa}.

\phead{Parameter Sizes.}
The size of an LLM, indicated by the number of parameters, affects its capacity to learn and generalize. Larger models often have more expressive power but require more computational resources~\cite{Kaplan2020ScalingLF}.

\phead{Pre-training Objectives.}
NLP-based LLMs are trained on natural language text and are designed to perform a variety of language tasks, including question-answering, translation, summarization, and text generation. Code-based LLMs are specifically trained on programming languages and are tailored to understand and generate code~\cite{CodeBERT, GraphCodeBERT}. However, some code LLMs, such as CodeLlama, are fine-tuned from LLaMA2, a NLP-based language model, rather than being trained exclusively on code. This fine-tuning process adapts the base model to better handle code-specific tasks~\cite{CodeLlama}.
% \zeyang{some code LLMs are code-fine-tuned from language models instead of only trained on code. such as codellama is fine-tuned from llama2. https://arxiv.org/pdf/2308.12950}

\subsubsection{\textbf{Learning Paradigms of LLMs}}

LLMs are highly adaptable due to their ability to leverage transfer learning, which enables them to apply pre-trained knowledge across various tasks. However, they may still face difficulties with certain tasks because of limited domain knowledge. To further tailor LLMs for more context-specific tasks, two key learning paradigms are commonly employed:
% \peter{1) you should say adapt instead of all the discussion on fine-tuning here}LLMs benefits from transfer learning, a process which involves taking an LLM that has been pre-trained on one corpus of text data and then fine-tuning it for a specific ``downstream'' task, such as text classification or text generation, by updating the model’s parameters with task-specific data. Transfer learning allows LLMs to be fine-tuned for specific tasks with much smaller amounts of task-specific data than would be required if the model were trained from scratch. This greatly reduces the amount of time and resources needed to train LLMs. For example, CodeBERT~\cite{CodeBERT}, codebert-java~\cite{codebert-java}, and GraphCodeBERT~\cite{GraphCodeBERT} have been trained specifically to understand and generate code. 

\phead{In-Context Learning.}
In-Context Learning (ICL) is a method where an LLM is provided with a prompt that includes detailed instructions, task-specific demonstrations, or both. This enables the model to adjust its responses to the new task based on the provided context, without altering its underlying parameters. By leveraging its pre-trained knowledge, the model generates responses based on the provided context~\cite{brown2020languagemodelsfewshotlearners, wang2020generalizingexamplessurveyfewshot}. 
However, ICL has limitations. The model's context size limits the number of examples that can be included, which may reduce its effectiveness. Additionally, processing multiple examples can increase computational and financial costs due to increased input tokens~\cite{Kaplan2020ScalingLF}, leading to longer inference times.

\phead{Fine-tuning. }
In fine-tuning, a LLM undergoes a secondary training phase on a more specific dataset related to a particular task or domain~\cite{dodge2020finetuningpretrainedlanguagemodels, gao2021makingpretrainedlanguagemodels, Radford2018ImprovingLU}. This process involves adjusting the model's parameters based on the new data, which enhances the model's performance on tasks relevant to the fine-tuning dataset. However, fine-tuning LLMs can be time-consuming because it involves updating full parameters. To address this, recent techniques have been introduced that only adjust additional parameters, thereby accelerating the fine-tuning process. Techniques like LoRA (Low-Rank Adaptation) can be used during fine-tuning to efficiently adapt pre-trained models by introducing low-rank matrices into the network, which helps manage the computational cost while preserving model performance~\cite{LoRA}. Unlike in-context learning, which temporarily adjusts the model based on the input, fine-tuning changes the model's behavior and knowledge. %Fine-tuning is also not limited by context length and can handle a wider range of data, improving the model's ability to generalize across various tasks.

\subsubsection{\textbf{Privacy Concerns}}
% \zeyang{This privacy issue is from commercial LLMs. All LLMs we used in this study are open-sourced. So, we should mention that commercial LLMs.....  Thus, open-source LLMs are more secure and trustworthy for handling confidential data such as codes. The discussion here can be similar with \href{https://www.arxiv.org/pdf/2408.01585#page=2.30}
%While LLMs are effective at handling natural language and code, using 
LLM with sensitive data, such as code, raises serious privacy concerns, especially with commercial models like ChatGPT. For instance, Samsung recently banned the use of ChatGPT among employees due to a sensitive code leak, which highlighted concerns over data security and interactions with proprietary information~\cite{forbesSamsungBans}. To prevent these privacy concerns, our study utilizes open-sourced LLMs. By choosing open-source models for local deployment, we ensure data privacy and adherence to strict protection standards, allowing us to explore the effective integration of LLMs with in-house code while safeguarding sensitive information and maintaining data security.

% \textbf{(2) Comparison of Traditional vs. LLM-Based Approaches. } Understanding how traditional methods, such as deep learning, as compare to LLM-based approaches in software engineering tasks is still needed.

% \zeyang{why we are saying "we address the challenges" here, we did not introduce many challenges in Section 2.1. We said "Despite these advancements, several areas remain unclear and
% warrant further investigation:" before 2.1.1. I feel it's better to say "we investigate how to leverage LLMs on log level prediction and compare to traditional approaches". }

By focusing on log level suggestion, our study explores under-examined areas in the application of LLMs, including challenges related to model fine-tuning, comparisons between traditional and LLM-based approaches, and privacy concerns.

\section{STUDY DESIGN}
\label{sec:datasrc}

% \peter{don't say this is a framework. Our results are worse than prior work. We should say this is an empirical study that compares in-context learning and fine-tuning using various LLMs for a specific code related task - log level suggestion. We also want to compare with traditional approaches to see if LLM-based is better or worse. One of the take-home is, LLM-based is not better than the traditional approach.}

% \begin{figure}[htb!]
%     \centering
%     \vspace{-0.2cm}
%     \includegraphics[width=0.95\linewidth]{figures/Log_prompt_StudyDesign.pdf}
%     \caption{An overview of our study design: Prompts containing processed source code and log messages are input into the LLM to examine the impact of various learning paradigms. The outcomes are subsequently collected and assessed.}
%     \label{fig:studydesign}
% \end{figure}
% Figure~\ref{fig:studydesign} depicts the process of our study design. The subsequent subsections provide a detailed explanation of each step.

\begin{figure*}[htb!]
    \centering
    % \vspace{-0.2cm}
    \includegraphics[width=0.95\linewidth]{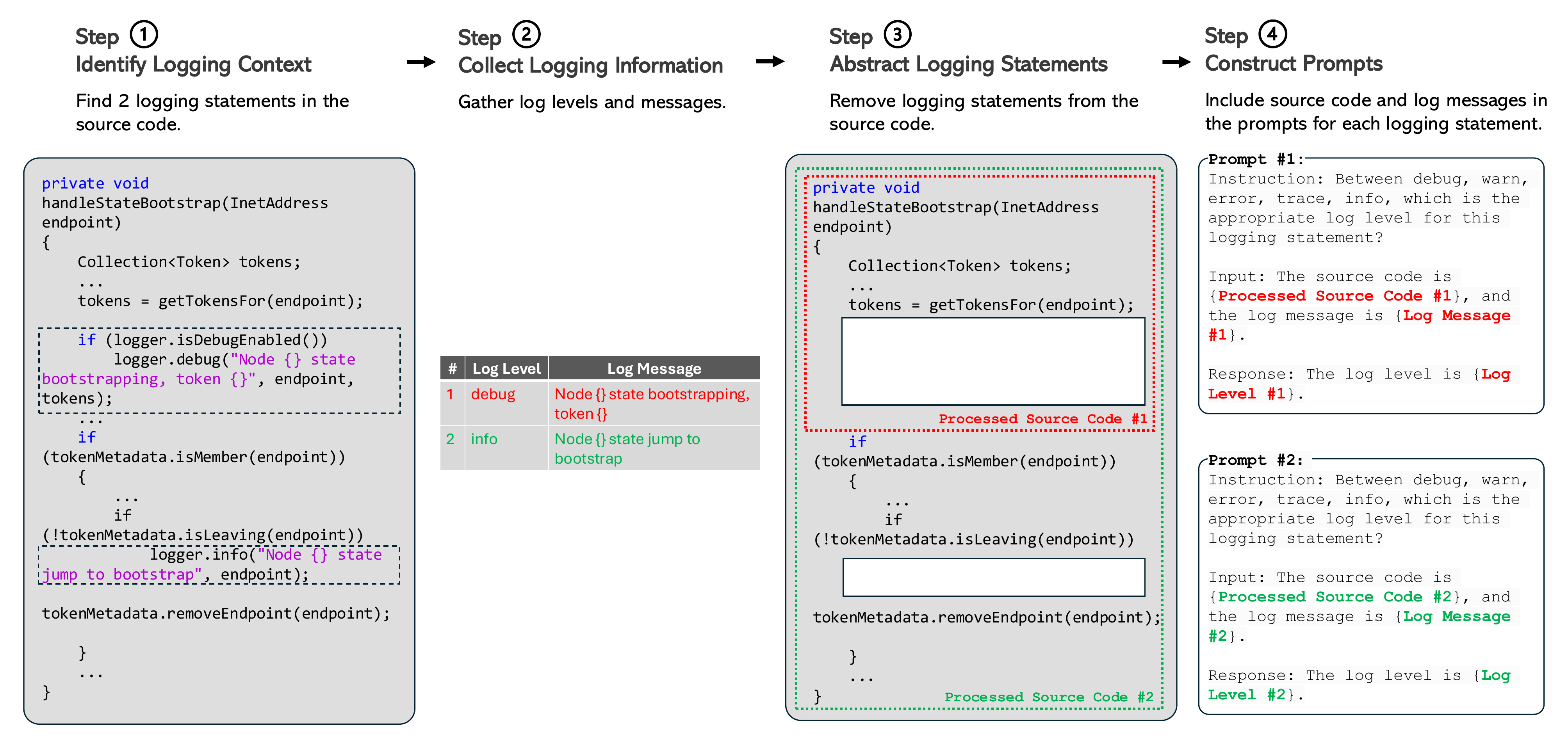}
    \caption{Outline of the data processing stage: \picircle{1} \textbf{Identify Logging Context:} Locate methods with logging statements, \picircle{2} \textbf{Collect Logging Information:} Determine the log levels and messages, \picircle{3} \textbf{Abstract Logging Statements:} Remove logging statements and log guards from the source code, \picircle{4} \textbf{Construct Prompts:} Create prompts for the LLM using the gathered information.}
    \label{fig.overview}
% \vspace{-10pt}  
\end{figure*}

\begin{table}[]
\caption{An overview of the log level distribution across nine large scale systems.}
\centering
\resizebox{\columnwidth}{!} {
\begin{tabular}{@{}l|lllrrrrrr@{}}
System & Version & LOC & NOL & \multicolumn{1}{l}{Trace} & \multicolumn{1}{l}{Debug} & \multicolumn{1}{l}{Info} & \multicolumn{1}{l}{Warn} & \multicolumn{1}{l}{Error} & \multicolumn{1}{l}{Fatal} \\ \midrule
Cassandra & 3.11.4 & 432K & 1.3K & 16.7\% & 10.9\% & 15.8\% & 16.8\% & 39.8\% & 0.0\% \\
ElasticSearch & 7.4.0 & 1.50M & 2.5K & 28.5\% & 32.4\% & 10.0\% & 19.2\% & 9.9\% & 0.0\% \\
Flink & 1.8.2 & 177K & 2.5K & 1.0\% & 30.8\% & 26.6\% & 23.7\% & 17.9\% & 0.0\% \\
HBase & 2.2.1 & 1.26M & 5.5K & 7.4\% & 17.3\% & 17.1\% & 24.4\% & 33.8\% & 0.0\% \\
JMeter & 5.3.0 & 143K & 1.9K & 0.7\% & 29.9\% & 16.9\% & 26.5\% & 26.0\% & 0.0\% \\
Kafka & 2.3.0 & 267K & 1.5K & 12.9\% & 28.5\% & 20.4\% & 15.3\% & 22.9\% & 0.0\% \\
Karaf & 4.2.9 & 133K & 0.8K & 0.9\% & 21.9\% & 23.1\% & 30.0\% & 23.6\% & 0.5\% \\
Wicket & 8.6.1 & 216K & 0.4K & 2.2\% & 39.3\% & 7.6\% & 28.5\% & 22.4\% & 0.0\% \\
Zookeeper & 3.5.6 & 97K & 1.2K & 2.2\% & 18.3\% & 19.3\% & 35.3\% & 24.9\% & 0.0\% \\ \midrule
Average & —— & 469K & 2.0K & 8.0\% & 25.5\% & 17.5\% & 24.5\% & 24.4\% & 0.1\% 
\end{tabular}
}
\label{table:logleveldistribution}
% \vspace{-8pt}  
\end{table}

\subsection{Studied Dataset}
\label{sec:studieddataset}
% \zeyang{I think we need also discuss these systems were studied by prior works.}
\phead{Overview.}
We conduct the study on nine large-scale open-source Java systems (Table~\ref{table:logleveldistribution}). We chose these systems because they are actively maintained by the Apache Software Foundation, well-documented, and cover a variety of domains from database systems to search engines. They vary in size, with Lines of Code (LOC) ranging from 97K to 1.5M and Number of Logging Statements (NOL) ranging from 0.4K to 5.5K. 
%Known for their high-quality logging code and adherence to robust logging practices,
These systems have also been widely used in prior research on logging~\cite{DeepLV, TeLL, Li2020WhereSW, DLFinder}.

We analyzed the logging statements across all nine systems. % previously examined by Li et al.~\cite{DeepLV}. 
Table~\ref{table:logleveldistribution} shows the log level distribution across these different systems. We observed that the primary use of logging statements is to highlight potential issues during system execution, with a notable distribution among the \textit{debug}, \textit{info}, \textit{warn}, and \textit{error} levels. For example, 25.5\% of the logging statements belong to the \textit{debug} level, 24.5\% to the \textit{warn} level, 24.4\% to the \textit{error} level, and 17.5\% to the \textit{info} level. However, we observed that \textit{fatal} log levels are present in only one of the nine systems studied (i.e., Karaf), accounting for less than 1\% of the total logging statements. As the \textit{fatal} level is considered obsolete in modern logging frameworks due to its similarity to the \textit{error} level, as stated in the SLF4J official documentation~\cite{slf4jfaq}, we excluded it from our study. Moreover, the \textit{trace} level has a lower occurrence: an average of 8.0\% of logging statements. Notably, projects such as Flink, JMeter, Karaf, Wicket, and Zookeeper exhibit a negligible number of \textit{trace}-level logging statements, ranging from 0.7\% to 2.2\%. According to the SLF4J official documentation, the \textit{trace} level shares a similar semantic meaning with the \textit{debug} level, which may explain the lower occurrence of logging statements at the \textit{trace} level. In summary, \textit{debug}, \textit{warn}, and \textit{error} collectively constitute 75\% of the logging statements. %\zeyang{I feel a better summary is to say the distribution of log levels is not even, which makes it challenging to find the correct log level. If we say that some levels have more distribution, it will seem that log level suggestion can be correct easily if we just try to predict the level to these frequent levels.}

The uneven distribution of log levels presents a significant challenge in achieving accurate log level suggestion. The disparity in frequency among different log levels means that some levels are more common than others, which implies that suggesting these more frequent levels might be simpler. However, this approach fails to account for the challenge of accurately identifying less frequent but still crucial log levels. Therefore, while it may appear that focusing on the most common log levels could improve suggestion accuracy, addressing the imbalance and ensuring accurate suggestions across all levels remains a complex and crucial task for effective log level suggestion.

\phead{Data Collection.}
Figure~\ref{fig.overview} illustrates the structure of our data preparation process for the log level suggestion. It consists of four key components: \circled{1} Identify Logging Context, \circled{2} Collect Logging Information, \circled{3} Abstract Logging Statement, and \circled{4} Construct Prompts. 

\phead{\circled{1} Identify Logging Context.}
We used static analysis to locate logging statements by traversing the abstract syntax tree to find method invocations using widely used logging libraries such as Log4j and SLF4J. This approach enables us to identify the context in which logging occurs by retrieving the calling method for each method containing a logging statement. Understanding the calling method is crucial because it helps us trace the source of the logging events, providing insight into the broader execution context and potentially identifying patterns or issues related to logging practices. 

\phead{\circled{2} Collect Logging Information.}
After identifying methods with logging statements, we extract and parse these statements to gather relevant information. Specifically, we collect (1) the log message and (2) the verbosity level. In our example, there are two logging statements with different log level identified in the method \texttt{handleStateBootstrap}. In total, we retrieved 17.6K logging statements from the nine systems.

% ~\djk{explain more about how we parsed this}
% ~\djk{What kind of block level do we care.}

% ~\djk{also talk about what if there is two logging statements, we breakdown this into two...  We need to make this part more interesting.}

\phead{\circled{3} Abstract Logging Statements.}
Our goal is to abstract logging statements by not only removing the logging statements but also excluding any code from basic blocks that follow the one containing each logging statement. For each logging statement, we extract the source code from the start of the method up to the end of the basic block where the logging statement is found. This approach ensures we consider only the relevant part of the code, as basic blocks provide a continuous sequence of statements without branching. As shown in Step 3 of Figure~\ref{fig.overview}, Log Statement \#2 is associated with a longer segment of code because it is located further down in the method, whereas Log Statement \#1 is in a shorter block.

In addition to removing logging statements, we also eliminate log guards, conditions that determine whether a log message should be processed based on current logging settings. We exclude all AST nodes related to log guards before sending the code context to LLMs, a practice that prevents data leakage and aligns with previous work~\cite{DeepLV, TeLL}.

\phead{\circled{4} Construct Prompts.} 
Prior studies~\cite{DeepLV, TeLL} have investigated various feature combinations for suggesting log levels. For instance, DeepLV~\cite{DeepLV} combines the syntactic context of the logging statement with the log message content, while TeLL~\cite{TeLL} integrates multi-level block information with log messages. Inspired by these methods, our study leverages the strengths of combining diverse information sources. As depicted in Step 4 of Figure~\ref{fig.overview}, we have developed a prompt template that incorporates two key features for input into the LLMs: (1) Processed Source Code and (2) Log Message.

\begin{table}
\centering
\caption{Language models used in our study.}
\resizebox{1.0\columnwidth}{!}{%
\begin{tabular}{c c l l}
\hline            
\textbf{Task} & \textbf{Pre-training} & \textbf{Model Name} & \textbf{Parameter}\\
\textbf{Type} & \textbf{Objective} &  & \textbf{Size} \\
\hline
\multirow{8}{*}{Fill-Mask} & \multirow{4}{*}{NLP-based} & BERT base & 110M \\
& & BERT large & 336M \\
& & RoBERTa base & 125M \\
& & RoBERTa large & 355M \\
\cline{2-4}
\multirow{0}{*}{} & \multirow{4}{*}{Code-based} & CodeBERTa & 84M \\
& & CodeBERT & 125M \\
& & codebert-java & 125M \\
& & GraphCodeBERT & 125M \\
\cline{2-4}
\hline
\multirow{4}{*}{Text Generation} & \multirow{2}{*}{NLP-based} & LLaMA 2 7B & 7B \\
& & LLaMA 2 13B & 13B \\
\cline{2-4}
\multirow{0}{*}{} & \multirow{2}{*}{Code-based} & CodeLlama 7B & 7B \\
& & CodeLlama 13B & 13B \\
\cline{2-4}
\hline
\end{tabular}
  }
  \label{table:languagemodels}
% \vspace{-5pt}
\end{table}
\subsection{Selecting Large Language Models (LLMs)}
\label{sec:llms}
We selected 12 open-source Large Language Models, as enumerated in Table~\ref{table:languagemodels}, for our experimental purposes. These models encompassing both NLP-based and Code-based models, vary in size, ranging from 84 million to 13 billion parameters, and will all be executed using identical datasets in \ref{sec:studieddataset} to examine the influence of both model size and type on performance, eliminating the potential confounding factor of using diverse training/testing data.
%\zhenhao{NLP-based (Natural Language Processing Based) vs. NL-based (Natural Language Based), maybe use NL-based throughout.}
%\yw{NLP and NL based sounds very much alike, can i keep the code based instead? }

\phead{Fill-Mask Models.}
NLP-based Language Models like BERT base, BERT large, RoBERTa base, and RoBERTa large are chosen for examination to assess whether programming statements can effectively extract semantic information. Specifically, these Language Models undergo pretraining using the Masked Language Modeling (MLM) objective. This involves randomly masking 15\% of the words in a sentence, subsequently processing the entire masked sentence through the model, and predicting the masked word~\cite{BERT}.
For code-based models, we opted for CodeBERT and CodeBERTa as they are pre-trained on data from various programming languages. %models for programming languages, which is a multi-programming-lingual model pre-trained on NL-PL pairs in 6 programming languages \cite{CodeBERT}. Our choice extended to CodeBERTa by HuggingFace, which underwent training on open-source projects from GitHub%, focusing on Go, Java, JavaScript, PHP, Python, and Ruby\cite{CodeBERTa-small-v1}. 
%It's important to note that both CodeBERT and CodeBERTa are initialized with RoBERTa-base\cite{CodeBERT, codebert-java}. 
We also consider, codebert-java, a specialized CodeBERTa model trained exclusively on Java code, shares the identical model architecture and size (125M parameters) due to their common foundation in the masked-language-modeling task~\cite{codebert-java}.

\phead{Text Generation Models.} We selected LLaMA 2 models as they provide a benchmark for evaluating large-scale, general-purpose language understanding and generation~\cite{LLaMA}.  While the CodeLlama models, specifically fine-tuned from LLaMA 2 models for coding tasks, offer insights into how specialized training affects performance in code-related applications. For both LLaMA 2 and CodeLlama, we selected 7B and 13B versions from both LLMs to compare and assess the impact of parameter size on accuracy in log level suggestion and investigate how these models perform under In-Context Learning and fine-tuning scenarios.

\subsection{Sampling Few Shot Data}
\label{Sampling}
Prior studies~\cite{Gong2018DiversityIM, Song2016CBrainAD, Ma_2024} have shown that using diverse training data improves the performance and generalization of deep learning models. Based on this, we select few shot samples from the target system to ensure we cover each available log levels. For example, for a 5-shot sample, we randomly select one logging sample from each level: debug, warn, error, trace, and info. For a 30-shot sample, we choose six logging samples from each log level. Random selection helps to avoid bias and ensures that our samples represent a wide range of scenarios, ensuring the diversity of the samples.

To explore how the number of shots affects LLM performance, we selected samples with 5, 10, 20, and 30 shots. These samples were then utilized in two different learning paradigms:
% We perform log level suggestion leveraging LLMs as indicated in Table~\ref{table:languagemodels}. For each LLM, we compare two prompting techniques for predicting log levels: (1) in-context learning and (2) fine-tuning. For in-context learning, we follow the prompting template from Figure~\ref{fig:template1}, choosing 5-shot examples, each sample representing one of the five log levels, along with the code context. Similarly, for fine-tuning, we utilize the same 5-shot examples used in the in-context learning settings for a direct comparison. However, for fine-tuning, we also leverage more examples to analyze how log-level suggestion improves as we train with larger datasets. We use 10, 20, and 30 logging statements, containing 2, 4, and 6 logging statements from each of the five log levels, respectively. 

\phead{In-Context Learning.}
Given the token limitations of the prompt, we concentrate on using 5-shot samples for In-Context Learning (ICL). In our experiment, each prompt includes five samples, with each sample containing the processed source code and log message as input, and the corresponding log level of a logging statement as the output. The goal is to assess if the LLM can accurately match log levels with functions and logging statements. We also evaluate ICL's effectiveness using a 0-shot prompt to determine how well the model can understand instructions without any prior examples.

\phead{Fine-tuning.}
To compare the performance of In-Context Learning and fine-tuning, we use the 5-shot samples employed in the ICL prompts for fine-tuning the LLMs. This comparison enables us to evaluate the differences in model performance between these two approaches: adapting to tasks with in-context examples versus updating the model's parameters through fine-tuning. Our analysis aims to determine which method provides superior accuracy for log level suggestion. 
Additionally, to assess the impact of the number of shots on LLM performance, we fine-tuned the models using 10, 20, and 30 logging samples, following the prompt template illustrated in Step 4 of Figure~\ref{fig.overview}. 
To ensure a fair comparison with prior works~\cite{DeepLV, TeLL}, we follow these studies by also fine-tuning the LLMs using 60\% of the data as a training dataset. We apply stratified random sampling~\cite{stratified} to divide the dataset into 60\% of the input data for training, 20\% for validation, and 20\% for testing. This approach maintains the same amount of data used in the training, as well as ensures distribution of log levels across all sampled datasets as in the original data.

\subsection{Evaluation Metrics}
A ``verbalizer" is a method for converting abstract labels or tokens into specific, interpretable terms. For Fill-Mask LLMs, log levels are used as tokens for verbalizers to provide clear mappings for suggestion tasks, whereas Text Generation LLMs do not require explicit verbalizers, as these models can inherently identify and apply suitable verbalizations for log levels. Unlike Fill-Mask LLMs, which focus on completing or predicting specific words within a given context, Text Generation LLMs are designed to produce coherent and contextually appropriate text sequences that extend beyond single words. To handle this, we use post-processing techniques to extract the relevant log level from their outputs, which aligning with methods used in related studies~\cite{Mastropaolo2022UsingDL}. 
Following prior studies~\cite{DeepLV,TeLL}, we use Accuracy, Area Under the Curve (AUC), and Average Ordinal Distance Score (AOD), to assess the performance of LLMs in suggesting log levels.

\phead{Accuracy.} The accuracy metric, widely employed in previous multi-class classification studies, as evidenced by investigations conducted by~\cite{8840982, DeepLV, UniLog, Mastropaolo2022UsingDL, TeLL}, assesses the proportion of correctly suggested log levels relative to the total number of log levels. 
We define accuracy as the percentage of correctly suggested log levels among all the results obtained from the suggestion process. A higher accuracy indicates a model's proficiency in accurately recommending log levels for a larger number of logging statements.

% ~\zeyang{you used "block" here instead of "level", it means if there are 2 loggings in one method, if one correct and one wrong. this method will be considered 1 wrong case, right? Or there is no cases that 2 loggings in one method? In fig~\ref{fig.overview}, you showed one method that contains 2 logging statements. When you send this method to LLM, you predict both of the log levels at the same time? } 

%\zhenhao{I think no need to present the formula of accuracy, to save some space.}
%The calculation of accuracy follows the formula:

%\begin{align}
%Accuracy = \frac{Number\ of\ Correct\ suggestions}{Total\ Number\ of\ suggestions}
%\end{align}

\phead{Area Under the Curve (AUC).}
The AUC (Area Under the Curve) measures a model’s ability to discriminate between different classes, based on the ROC (Receiver Operating Characteristic) curve, which plots the true positive rate against the false positive rate. AUC values range from 0 to 1, with a higher value indicating better discriminative ability. An AUC below 0.5 suggests performance no better than random guessing. In this study, we use the multiclass AUC definition as established by Hand et al.~\cite{Hand2001ASG}, following the approach used in prior researches~\cite{Li2016, DeepLV, TeLL}. In both previous research and our study, this metric shows a model's proficiency in distinguishing between various log levels. 

\phead{Average Ordinal Distance Score (AOD).}
%This metric, as presented by DeepLV~\cite{DeepLV}, addresses the limitations observed in the aforementioned accuracy and AUC metrics. While accuracy and AUC treat log levels as independent classes, 
It assesses the proximity between the actual log level and the suggested log level for each logging statement~\cite{DeepLV}. Each log level is assigned a numerical value, and the AOD is computed using the following formula:
\begin{align}
AOD = \frac{\sum_{i=1}^{N}\left(1 - \dfrac{Dis(a_{i},s_{i})}{MaxDis(a_{i})}\right)}{N}
\end{align}
where $N$ represents the total number of suggestions. For each logging statement and its suggested log level, {\em Dis(a, s)} is the distance between the {\em actual} log level $a_{i}$ and the {\em suggested} log level $s_{i}$ (e.g., the distance between {\em error} and {\em info} is 2). The maximum possible distance of the actual log level ai is denoted by {\em MaxDis(a)}. The resulting AOD value ranges from 0 to 1, with a higher value indicating that the suggested log level is closer to the actual log level.

%Following the introduction of AOD, various papers~\cite{TeLL, Li2023ExploringTE} have adopted and employed this metric in their respective analyses on log levels. Therefore, we also utilize AOD in our study to ensure consistency with established methods and to leverage its proven effectiveness in evaluating log level suggestion.

\phead{Environment and Implementation.}
Our experiments were conducted on a server with an NVIDIA Tesla V100 GPU using CUDA 12.2.2. 
We use the OpenPrompt framework~\cite{openprompt2021} to fine-tune Fill-Mask LLMs by masking words in sentences and suggesting appropriate log levels. For Text Generation models, we fine-tune using LoRA~\cite{LoRA}, applying a maximum learning rate of 5e-4, the AdamW~\cite{AdamW} optimizer, and a linear learning rate decay schedule.
LLMs exhibit inherent randomness during the inference process. To ensure consistent output for the same input, we set the temperature parameter to 0.

% \section{EXPERIMENT SETUP AND IMPLEMENTATION}
% \label{sec:experiment}

\section{Results}
\label{sec:results}

In this section, we present our study results by answering the research questions (RQs).

\begin{table*}
  \caption{Comparison of accuracy, AUC, and AOD of LLMs across in-context learning and fine-tuning with varying few-shot sizes. The \underline{highest} value for Fill-Mask LLMs is denoted in \textcolor{blue}{\textbf{blue}}, and the \underline{highest} value for Text Generation LLMs is in \textcolor{red}{\textbf{red}}.}
\resizebox{2.05\columnwidth}{!}{%
\begin{tabular}{l|c|c|c|c|c|c|c|c|c|c|c|c|c|c|c|c|c|c|c|c|c}\toprule
\multirow{3}{*}{LLMs} &\multicolumn{6}{c|}{\textbf{In-Context Learning }} &\multicolumn{15}{c}{\textbf{Fine-tuning}} \\\cline{2-22}
&\multicolumn{3}{c|}{0 shots} &\multicolumn{3}{c|}{5 shots} &\multicolumn{3}{c|}{5 shots} &\multicolumn{3}{c|}{10 shots} &\multicolumn{3}{c|}{20 shots} &\multicolumn{3}{c|}{30 shots} &\multicolumn{3}{c}{60\% shots} \\  
\cline{2-22}           
& Acc. & AUC & AOD & Acc. & AUC & AOD & Acc. & AUC & AOD & Acc. & AUC & AOD & Acc. & AUC & AOD & Acc. & AUC & AOD & Acc. & AUC & AOD \\
\cline{1-22}                       

\textbf{Fill-Mask} &  &  &  &  &  &  &  &  &  &  &  &  &  &  &  &  &  &  &  &  &  \\		
BERT base & 18.67 & 53.12 & 57.65 & 22.58 & 57.94 & 57.58 & 29.30 & 59.69 & 54.73 & 31.43 & 60.78 & 54.16 & 35.64 & 64.00 & 61.06 & 38.38 & 65.51 & 63.19 & 60.77 & 81.08 & 79.03 \\		
BERT large & 18.79 & 53.37 & 58.24 & 21.84 & 57.97 & 62.13 & 31.47 & 62.00 & 58.75 & 36.03 & 64.30 & 59.81 & 39.54 & 67.03 & 65.23 & 38.91 & 66.84 & 65.07 & 61.4 & 81.48 & 79.86 \\		
RoBERTa base & 18.47 & 53.10 & 57.83 & 19.93 & 54.32 & 53.55 & 29.64 & 59.50 & 57.81 & 35.01 & 64.31 & 62.32 & 38.62 & 65.99 & 63.9 & 41.91 & 68.77 & 67.36 & 62.74 & 82.23 & 80.18 \\		
RoBERTa large & \textcolor{blue}{\textbf{27.40}} & 54.10 & 57.35 & 21.50 & 52.58 & 50.01 & 30.57 & 60.41 & \textcolor{blue}{\textbf{61.03}} & 36.31 & 64.84 & 63.67 & 40.58 & 65.97 & 65.48 & 45.76 & 70.64 & 70.26 & 63.90 & 82.78 & 81.31 \\		
CodeBERTa & 23.12 & 56.09 & 59.21 & 27.93 & 61.32 & 63.50 & 28.93 & 59.34 & 56.30 & 29.98 & 59.36 & 55.96 & 34.53 & 61.80 & 59.34 & 38.05 & 64.51 & 62.69 & 60.73 & 80.54 & 78.38 \\		
CodeBERT & 23.07 & \textcolor{blue}{\textbf{57.03}} & 68.65 & \textcolor{blue}{\textbf{31.76}} & \textcolor{blue}{\textbf{65.20}} & 68.83 & 31.55 & 62.86 & 56.63 & 36.75 & 66.76 & 62.20 & 42.47 & 71.24 & 69.33 & 46.46 & 72.78 & 71.02 & 63.88 & 82.88 & 81.08 \\		
codebert-java & 23.41 & 56.66 & \textcolor{blue}{\textbf{}}68.98 & 26.46 & 58.41 & \textcolor{blue}{\textbf{69.50}} & \textcolor{blue}{\textbf{35.48}} & 64.52 & 58.75 & 38.57 & 65.98 & 61.48 & 45.04 & 71.22 & 69.86 & 47.16 & 72.49 & 71.54 & 63.42 & 82.16 & 80.60 \\		
GraphCodeBERT & 22.56 & 55.99 & 68.71 & 30.01 & 66.48 & 68.02 & 34.07 & \textcolor{blue}{\textbf{65.00}} & 58.55 & \textcolor{blue}{\textbf{40.41}} & \textcolor{blue}{\textbf{69.51}} & \textcolor{blue}{\textbf{66.53}} & \textcolor{blue}{\textbf{45.60}} & \textcolor{blue}{\textbf{72.50}} & \textcolor{blue}{\textbf{71.36}} & \textcolor{blue}{\textbf{48.61}} & \textcolor{blue}{\textbf{74.02}} & \textcolor{blue}{\textbf{71.76}} & \textcolor{blue}{\textbf{64.81}} & \textcolor{blue}{\textbf{83.20}} & \textcolor{blue}{\textbf{81.28}}  \\		
 &  &  &  &  &  &  &  &  &  &  &  &  &  &  &  &  &  &  &  &  &  \\
\textbf{Text Generation} &  &  &  &  &  &  &  &  &  &  &  &  &  &  &  &  &  &  &  &  &  \\
LLaMA 2 7B & 29.60 & 60.18 & 66.45 & 27.73 & 57.88 & 64.7 & 27.77 & 62.37 & 58.52 & 29.39 & 59.73 & 57.18 & 33.5 & 66.42 & 68.2 & 34.51 & 64.61 & 68.28 & 38.14 & 60.9 & 66.51  \\
LLaMA 2 13B & 31.56 & 64.44 & 70.85 & 24.99 & 56.45 & 67.35 & 27.98 & 61.16 & 58.27 & 33.85 & 66.54 & 63.74 & 37.82 & 69.44 & 68.42 & 41.22 & 71.55 & 71.84 & 55.13 & 78.66 & 79.83  \\
CodeLlama 7B & 34.22 & 64.16 & 72.09 & \textcolor{red}{\textbf{44.83}} & \textcolor{red}{\textbf{75.53}} & \textcolor{red}{\textbf{78.87}} & 33.69 & 64.64 & 66.93 & 37.68 & 68.87 & 65.19 & 42.37 & 73.09 & 72.69 & 42.75 & 73.21 & 70.79 & \textcolor{red}{\textbf{58.33}} & \textcolor{red}{\textbf{80.45}} & \textcolor{red}{\textbf{81.56}}  \\
CodeLlama 13B & \textcolor{red}{\textbf{42.40}} & \textcolor{red}{\textbf{74.01}} & \textcolor{red}{\textbf{76.22}} & 39.58 & 69.98 & 69.98 & \textcolor{red}{\textbf{41.92}} & \textcolor{red}{\textbf{73.73}} & \textcolor{red}{\textbf{74.78}} & \textcolor{red}{\textbf{41.53}} & \textcolor{red}{\textbf{72.45}} & \textcolor{red}{\textbf{70.77}} & \textcolor{red}{\textbf{44.00}} & \textcolor{red}{\textbf{75.41}} & \textcolor{red}{\textbf{74.89}} & \textcolor{red}{\textbf{43.78}} & \textcolor{red}{\textbf{75.06}} & \textcolor{red}{\textbf{75.18}} & 45.53 & 75.75 & 77.31  \\
\bottomrule         
\end{tabular}}

% \vspace{-5pt}  
\label{table:RQ1}
\end{table*}

\begin{table}
\centering
\caption{Comparison between our LLM-based log level suggestion: FM (Fill-Mask LLM) and TG (Text Generation LLM), and two state-of-the-arts deep learning-based techniques: DeepLV~\cite{DeepLV} and TeLL~\cite{TeLL}. }
\resizebox{1.00\columnwidth}{!}{%
 \tabcolsep=2pt
\begin{tabular}{lcccccccccccc}
   
\toprule
\multirow{2}{*}{Project} & \multicolumn{4}{c}{Accuracy} & \multicolumn{4}{c}{AUC} & \multicolumn{4}{c}{AOD} \\
\cmidrule(lr){2-5} \cmidrule(lr){6-9} \cmidrule(lr){10-13}
 & DeepLV & TeLL & FM & TG & DeepLV & TeLL & FM & TG & DeepLV & TeLL & FM & TG \\
\midrule
Cassandra & 60.6 & 63.5 & 62.2 & 41.7 & 84.2 & 88.4 & 80.0 & 66.4 & 80.5 & 81.2 & 78.4 & 73.1 \\
ElasticSearch & 57.7 & 70.3 & 55.8 & 63.9 & 81.3 & 90.5 & 77.1 & 82.9 & 80.2 & 84.1 & 77.4 & 82.9 \\
Flink & 65.2 & 72.9 & 72.0 & 75.2 & 85.1 & 92.5 & 87.1 & 88.6 & 83.8 & 86.3 & 85.2 & 87.3 \\
Hbase & 60.3 & 70.7 & 64.5 & 67.3 & 84.2 & 92.1 & 83.5 & 85.3 & 81.7 & 87.3 & 80.4 & 82.7 \\
Jmeter & 62.3 & 73.7 & 69.5 & 70.2 & 83.9 & 92.1 & 85.1 & 86.0 & 80.9 & 87.2 & 83.7 & 86.7 \\
Kafka & 51.8 & 64.2 & 60.6 & 53.5 & 79.5 & 88.8 & 81.8 & 81.8 & 77.5 & 81.2 & 80.9 & 79.8 \\
Karaf & 67.2 & 75.0 & 67.9 & 58.0 & 85.6 & 90.8 & 85.6 & 86.5 & 81.6 & 86.7 & 83.6 & 84.6 \\
Wicket & 63.8 & 74.4 & 63.7 & 50.0 & 85.0 & 89.9 & 82.2 & 68.7 & 79.3 & 85.6 & 78.8 & 78.5 \\
Zookeeper & 60.9 & 74.6 & 67.0 & 45.0 & 84.8 & 92.4 & 86.5 & 77.8 & 82.0 & 88.7 & 83.0 & 78.4 \\
\midrule
\textit{Average} & 61.1 & 71.0 & 64.8 & 58.3 & 83.7 & 90.8 & 83.2 & 80.4 & 80.8 & 85.4 & 81.3 & 81.6 \\
\bottomrule 
\end{tabular}%
}

\label{table:RQ1_comparison}
% \vspace{-10pt}  
\end{table}

% \begin{figure}
%         \vspace{-0.3cm}
%         \centering
%     \includegraphics[width=0.85\linewidth]{figures/RQ1_emptyfrequency.pdf}
%     \vspace{-0.3cm}
%       \caption{Invalid Log Level Predictions. Code LLaMA has less invalid log level prediction as compared to LLaMA 2.}
%       \label{figure:RQ1_emptyfrequency}
%       \vspace{-0.3cm}
%     \end{figure}

\subsection{RQ1: What is the accuracy of LLMs in suggesting log level? 
}

\phead{Motivation. } While LLMs have been used in prior studies to generate log statements~\cite{UniLog, Li2023ExploringTE, Mastropaolo2022UsingDL}, there is limited research on comprehensively suggesting log levels across diverse LLMs. Hence, we investigate (1) whether LLMs trained with different objectives (Fill-Mask and Text Generation Task) and different datasets (code snippets vs. natural language) vary in their effectiveness at suggesting log levels and (2) whether in-context learning or fine-tuning performs better in log-level suggestions. In particular, we answer the following three RQs to benchmark the capability of LLMs in log level suggestion:

\noindent \textbf{RQ1-A:} What is the effectiveness of LLMs in log level suggestion?

\noindent \textbf{RQ1-B:} How does fine-tuning or in-context learning impact LLMs?

\noindent \textbf{RQ1-C:} How does the performance of LLMs in log level suggestion compare to existing state-of-the-art?

\subsubsection{RQ1-A:  What is the effectiveness of LLMs in log level suggestion}
\mbox{} %   
    
\phead{Results.} %\textit{\textbf{LLMs with more parameters generally give better log level suggestion results within the same LLM, although the difference can be smaller, suggesting smaller models may be more cost effective due to higher performance. Moreover, we notice that the Fill-Mask Language Model performs better.}} 
\textit{\textbf{For log level suggestion tasks, while larger LLMs generally show better accuracy, smaller code-specific models like CodeLlama 7B can achieve comparable performance to their larger counterparts with lower resource consumption, making them a more efficient option.
}}Table~\ref{table:RQ1} shows the performance of log level suggestion across diverse fill-mask and Text Generation LLMs. We consistently observe that larger variant of LLMs (RoBERTa large, LLaMA 2 13B) which are trained with more parameters outperform their base counterparts. For instance, the accuracy of roberta-large surpasses that of roberta-base by 6.95\%. Similarly, both 13B versions of LLaMA 2 and CodeLlama exhibit slight but discernible improvements over their 7B counterparts, with 2.06\% and 0.88\% higher accuracy, respectively. However, the improvement in accuracy achieved by using larger variants of LLMs is not as substantial as the improvement gained from LLMs that have been trained with code. 
Interestingly, CodeLlama 7B performs comparably to CodeLlama 13B, despite having fewer parameters. Hence, for log level suggestion, the 7B model may be sufficient to achieve high performance with less resource consumption. 

\textit{\textbf{Choosing an LLM trained on relevant data is more crucial than selecting one trained with a larger number of parameters.}} Code-based LLM achieves superior performance (20\% to 40\% higher) compared to NLP-based LLMs, despite having smaller parameters. As expected, for both in-context learning and fine-tuning, LLMs trained on code achieve the highest accuracy, AUC, and AOD in both Fill-Mask and Text Generation tasks. Among fill-mask LLMs, the highest accuracy is achieved by GraphCodeBERT, while among text generation models, CodeLlama 7B leads. For instance, when comparing the accuracies of GraphCodeBERT and CodeLlama 7B, both achieve around 40\%, whereas NLP-based LLMs like BERT and Llama2 achieve slightly more than 20\% accuracy. 

\subsubsection{RQ1-B: How does In-Context Learning or fine-tuning impact LLMs} 
\mbox{} % 

\phead{Results. } \textit{\textbf{When using the same set of samples, 10 out of 12 LLMs perform better when these samples are used for fine-tuning rather than for prompt construction.}} Out of 8 Fill-Mask LLMs, 7 exhibited improved performance when the 5-shot samples were used for fine-tuning rather than for In-Context Learning. The exception was CodeBERT, which saw a slight accuracy decrease of 0.3\%. The other Fill-Mask LLMs demonstrated performance gains, with RoBERTa large showing the most significant improvement, up to 9.7\%. In contrast, among the Text Generation LLMs, only CodeLlama 7B experienced a minor decline in performance.

\textit{\textbf{Text Generation LLMs demonstrate superior performance out of the box, but they are difficult to fine-tune effectively.}} 
Although all LLMs showed potential for improvement through fine-tuning, the rate of improvement varied between Fill-Mask and Text Generation LLMs when fine-tuning with 10, 20, and 30 shots. Off-the-shelf Text Generation LLMs, with accuracies ranging from 29.60\% to 42.40\%, outperform Fill-Mask LLMs, which have accuracies between 18.47\% and 27.40\%. However, increasing the data to 60\% did not improve results for Text Generation LLMs. In contrast, Fill-Mask LLMs showed significant gains across all metrics, with accuracies nearly tripling.

\subsubsection{RQ1-C: How does the performance of LLMs in log level suggestion compare to existing state-of-the-art}
\mbox{} % 

% \zhenhao{I think the motivation for this sub-RQ below is not very necessary.}
%\phead{Motivation. } Both SOTA studies explored multiple variants: DeepLV~\cite{DeepLV} introduced DL(Syn) and DL(Msg), utilizing syntactic context and log message features separately, and DL(Comb), which combines both features. Meanwhile, TeLL~\cite{TeLL} introduced TeLL(Mul), using Multi-level Code Block Information for log level suggestion as the first variant, and TeLL(Comb), which combines Multi-level Code Block Information with log message as the second variant. Both studies demonstrated that combining these features yields the best performance. Therefore, we conduct a comparative analysis of the Comb versions in Table~\ref{table:RQ1_comparison}, along with the top-performing LLMs in both Fill-Mask and Text Generation types (GraphCodeBERT and CodeLlama 7B).

%Since both DeepLV and TeLL uses 60\% of the data to train, 20\% of the data to validate, and remaining 20\%, we split our data with the same ratio to ensure a fair comparison. We also apply the stratified random sampling~\cite{stratified} to maintain the distribution of log levels across all sampled datasets as in the original data. 

\phead{Results. } 
\textit{\textbf{Given the log message and source code of the method provided, fine-tuned Fill-Mask LLM achieves competitive performance with state-of-the-art models, with a margin under 3\%.}} As noted in RQ1-B, Text Generation LLMs performed poorly even after fine-tuning with 60\% of the input data. Specifically, CodeLlama 7B achieved 44\% accuracy, comparable to DeepLV models trained with log messages (42.0\%). On the other hand, Fill-Mask LLMs benefited significantly from fine-tuning, surpassing DeepLV in accuracy, AUC, and AOD metrics with GraphCodeBERT. Notably, accuracy improved by 7.5\%, though it still lags behind TeLL by 2.4\%. This demonstrates that a fine-tuned Fill-Mask LLM achieves competitive performance with state-of-the-art models when provided with both source code and log messages.

\phead{Discussions. }
\textit{\textbf{Text Generation LLMs, especially NLP-based LLMs, pose a high risk of producing invalid results, but this risk can be mitigated with the aid of fine-tuning.}}
We have noticed that Text Generation LLMs often rephrase questions or repeat instructions, leading to invalid suggestions. This issue, called ``hallucination''~\cite{rawte2023survey,shen2023chatgpt,yu2024fight}, refers to generating text that is nonsensical or strays from the original content. These hallucinations motivate us to investigate the frequency of unavailable results and to explore strategies for minimizing their occurrence.
 % \peter{not so clear how we got these numbers and what the numbers mean}
% \yw{There's no table for the number of invalid results because it was taking too much space, therefore we write it here.}
Among Text Generation LLMs, the CodeLlama 13B model has the lowest hallucination rate, with only a 0.19\% likelihood of producing an invalid output—about 2 out of 1,000 log level suggestions—despite being untrained and lacking example prompts. This shows that CodeLlama 13B performs the strongest in interpreting our prompts and generating appropriate responses. 

% Among all learning paradigms for CodeLlama 13B, the fine-tuning with 5 shots has the highest probability of returning an invalid output, at 2.62\%, while other paradigms show less than 1\%.
% Similarly to the results by CodeLlama 13B, CodeLlama 7B maintains a 5.73\% maximum  possibility in returning invalid results by fine-tuning with 5 shots. However, CodeLlama 7B has a 20.1\% chance of returning invalid input when the model is untrained and prompt does not contain an example, which is the opposite of CodeLlama 13B. However, the accuracy of CodeLlama 7B and CodeLlama 13B are both 34.22\% and 42.40\%, respectively, which means although 1 out of 5 suggested results are invalid, CodeLlama 7B has higher accuracy in log level suggestion.
Among all learning paradigms for CodeLlama 13B, fine-tuning with 5 shots has the highest probability of returning an invalid output at 2.62\%, while other paradigms show less than 1\%. Similarly, CodeLlama 7B reaches a 5.73\% possibility of returning invalid results with fine-tuning and 5 shots. However, CodeLlama 7B has a 20.1\% chance of returning invalid input when untrained and without examples in the prompt, opposite to CodeLlama 13B. Despite this, CodeLlama 7B and 13B have accuracies of 34.22\% and 42.40\%, respectively, meaning that although 1 in 5 suggested results are invalid, CodeLlama 7B achieves higher accuracy in log level suggestion.

As for LLaMA 2 variants, off-the-shelf LLaMA 7B almost have similar chances of returning invalid data, regardless of the samples provided in the prompt: ICL 0 shots have 25.41\% and ICL 5 shots have 24.30\% chance of returning an invalid response. Albeit the chances are still higher than Code LLaMA variants, trained LLaMA 2 are less likely to return invalid results benefit from the fine-tuning. However, LLaMA 2 13B has the highest possibility of getting confused by the long prompts and return invalid results 49.8\% of the time.

These observations point to the effectiveness of LLMs in different scenarios and underscore the importance of selecting the right model based on the context and requirements of the task. Enhanced training and fine-tuning strategies could mitigate the higher invalid output rates observed in some models, ultimately leading to more reliable and accurate performance in real-world applications.

\textit{\textbf{Insufficient training data may lead to monotonous log level predictions in Fill-Mask LLMs.}}
We noticed a trend indicating that, under the ICL 0 shots paradigm, Fill-Mask LLMs predominantly suggest either ``Info'' or ``Error'' log levels. As the amount of training data increases, these models gradually start incorporating a broader range of log levels in their suggestions. This trend underscores the impact of additional training data on enhancing the accuracy and flexibility of LLMs, particularly in adapting to a broader spectrum of log levels. This finding is crucial for improving the robustness and applicability of LLMs in real-world scenarios where diverse log level suggestions are required.

\rqbox{Fill-Mask models such as GraphCodeBERT, when fine-tuned, can outperform larger models in log level suggestions. Fine-tuning using task-specific data leads to substantial improvements in accuracy and makes these models competitive with state-of-the-art methods.}

\subsection{RQ2: How effective is including additional context in log level suggestion?}
\label{sec:RQ2}
%~\djk{reviewer might ask, why didn't we use this in the initial prompting?    }

% \peter{try to shorten this part}Prior studies~\cite{DeepLV, TeLL} compared the performance of log level suggestion deep learning frameworks by collecting two features: log messages and syntactic information. They found that combining both features yielded the best performance. For example, DeepLV~\cite{DeepLV} achieved an accuracy of 42.0\% when training their framework with only log messages, 54.0\% when training with only syntactic context, and 61.1\% when training with both features. TeLL~\cite{TeLL} achieved an accuracy of 71.0\% when using both log messages and multi-level block information, compared to 67.8\% when using only multi-level block information.
% Therefore, we used two features (source code and log message) in our initial prompting to have a fair comparison between LLMs and prior works. However, TELL~\cite{TeLL} has credited intra-block feature within their multi-level block information for their superior results. This prompts us to consider incorporating other contexts that may increase the accuracy of log level suggestion. 

\phead{Motivation. } Prior studies~\cite{DeepLV, TeLL} found that combining two features improves log level suggestion performance. DeepLV~\cite{DeepLV} achieved 42.0\% accuracy with log messages alone, 54.0\% with syntactic context, and 61.1\% with both. TeLL~\cite{TeLL} reached 71.0\% using log messages and multi-level block information, compared to 67.8\% with just the latter. In our comparison of LLMs and prior works, we initially included both source code and log messages. However, since TeLL~\cite{TeLL} credited intra-block features for their success, we are considering additional contexts to improve accuracy.

%Prior researches~\cite{DeepLV, TeLL} have compared the performance of their log level suggestion deep learning framework by collecting two features, using each of the features to test their framework with limited results, and concluded that a combination of two features is the best performance. For example,~\cite{DeepLV} collected the log messages and the syntactic information of a code block associated with the log message. 
%They further test the two steps: Training the deep learning framework with only log messages which achieved an average accuracy of 42.0\% and training the deep learning framework with only syntactic context only that yields 54.0\%. However, if the deep learning framework is trained with both log messages and  syntactic information, the accuracy has improved to 61.1\% as seen in Table~\ref{table:RQ1_comparison}. Similarly, TeLL~\cite{TeLL} achieved an accuracy of 71.0\% when using both log messages and multi-level block information, as compared to when 67.8\% when they use only multi-level block information to train their framework.

\phead{Approach.} The prompt string is tweaked to include the calling method as seen in the sample below.
\vspace{-3pt}
\begin{center}
\begin{mdframed}[nobreak=true, backgroundcolor=gray!20, linewidth=0pt, roundcorner=10pt, innerrightmargin=3pt, innerleftmargin=3pt, innertopmargin=3pt, innerbottommargin=3pt]
  % \begin{minipage}{1.0\textwidth}
    \begin{sloppypar}
\#\#\# Instruction: Between debug, warn, error, trace, info, which is the appropriate log level for this logging statement?

\noindent \#\#\# Input: The previous method is \textit{z}, the source code is \textit{x}, and the log message is \textit{y}.

\noindent \#\#\# Response: The log level is \textit{z}.
    \end{sloppypar}    
  % \end{minipage}
\end{mdframed}
\end{center}

\begin{table*}
\caption{Comparison of Fill-Mask Language Model (GraphCodeBERT) and Text Generation Language Model (CodeLlama 7B) with and without additional context.}
\centering
\resizebox{2.05\columnwidth}{!}{%
\begin{tabular}{l|c|c|c|c|c|c|c|c|c|c|c|c}\toprule
\multirow{3}{*}{Settings} &\multicolumn{6}{c|}{\textbf{Fill-Mask}}&\multicolumn{6}{c}{\textbf{Text Generation}} \\\cline{2-13}
&\multicolumn{2}{c|}{Acc.} &\multicolumn{2}{c|}{AUC} &\multicolumn{2}{c|}{AOD} &\multicolumn{2}{c|}{Acc.} &\multicolumn{2}{c|}{AUC} &\multicolumn{2}{c}{AOD} \\
\cline{2-13}
& W/O & With & W/O & With & W/O & With & W/O & With & W/O & With & W/O & With \\
\hline
ICL 0 shots & 38.80 & 38.88 (+0.08) & 70.39 & 68.18 (-2.21) & 68.39 & 67.20 (+0.21) & 40.38 & 37.61 (-2.77) & 67.59 & 65.47 (-2.12) & 71.34 & 70.45 (-0.89) \\
ICL 5 shots & 34.55 & 29.48 (-5.07) & 69.38 & 60.45 (-8.93) & 65.88 & 47.8 (+5.43) & 48.19 & 32.22 (-15.97) & 74.34 & 61.66 (-12.68) & 75.72 & 67.83 (-7.89) \\
FT 5 shots & 37.19 & 33.42 (-3.77) & 69.94 & 66.31 (-3.63) & 65.41 & 64.38 (-0.9) & 40.42 & 37.48 (-2.94) & 67.69 & 65.43 (-2.26) & 71.35 & 70.43 (-0.92) \\
FT 10 shots & 38.56 & 41.12 (+2.56) & 67.87 & 69.84 (+1.97) & 60.67 & 70.27 (-9.17) & 40.3 & 37.32 (-2.98) & 67.55 & 65.35 (-2.20) & 71.26 & 70.37 (-0.89) \\
FT 20 shots & 40.86 & 42.03 (+1.17) & 71.68 & 70.72 (-0.96) & 67.18 & 71.39 (-3.54) & 40.39 & 37.33 (-3.06) & 67.62 & 65.39 (-2.23) & 71.27 & 70.34 (-0.93) \\
FT 30 shots & 40.89 & 41.57 (+0.68) & 71.85 & 71.08 (-0.77) & 69.24 & 70.86 (-1.84) & 40.41 & 37.34 (-3.07) & 67.73 & 65.36 (-2.37) & 71.41 & 70.40 (-1.01) \\
FT 60\% shots & 67.3 & 62.88 (-4.42) & 83.43 & 81.90 (-1.53) & 82.99 & 80.78 (+1.09) & 40.43 & 37.61 (-2.82) & 67.80 & 65.57 (-2.23) & 71.44 & 70.53 (-0.91) \\
\bottomrule
\end{tabular}}
\label{table:RQ2}
\vspace{-15pt}  
\end{table*}

% \noindent 
\phead{Results.} 
\textit{\textbf{Overloading context negatively impacts LLM performance across learning paradigms.}} Table~\ref{table:RQ2} showed that including additional context will have a detrimental effect on the performance of LLMs in suggesting log level. Both Fill-Mask and Text Generation LLMs showed a slight decline in accuracy for each of the learning paradigms, especially in In-Context Learning 5 shots, where the results plummeted 15.97\% whereas other settings observed a decrease between 2.77\% to 3.07\%. This outcome conforms to the results in the work of He et al.~\cite{HeCharacterizing} and Shi et al.~\cite{shi2023largelanguagemodelseasily}, where incorporating more contexts cannot always guarantee a better result. 

\textit{\textbf{Short prompts decrease the risk of invalid outputs in log level suggestions using Text Generation LLMs.}}
% We examined how the inclusion of additional context affects the frequency of invalid outputs compared to instances where this extra information was not used. We found that longer prompts, which included calling methods as additional context, resulted in a higher rate of invalid outputs. This finding aligns with the observations of Schäfer et al.~\cite{schafer2023empiricalevaluationusinglarge} that adding details like the code of the calling method can confuse the model and produce incorrect results. Thus, our study highlights the advantages of using concise prompts to enhance the effectiveness of Text Generation LLMs in suggesting log levels.
We examined how including additional context affects the frequency of invalid outputs compared to cases without it. We found that longer prompts, including calling methods, led to a higher rate of invalid outputs. This finding is consistent with the research by Schäfer et al.~\cite{schafer2023empiricalevaluationusinglarge} and Liu et al.~\cite{liu2023lostmiddlelanguagemodels}, who noted that models struggle to robustly access and use information in long input contexts. In our case, incorporating details such as method call code can confuse the model, leading to invalid results and reduced accuracy. Our study highlights the advantages of concise prompts in improving Text Generation LLMs' effectiveness for log level suggestions.

\phead{Discussions.} 
\textit{\textbf{Concise Prompts for Enhancing LLM Performance in Log Level Suggestions.}} 
These findings suggest that overloading LLMs with excessive information can hinder their performance. By emphasizing concise prompts that focus on essential elements, such as log messages and source code, developers can improve the quality of the model's outputs. Our study highlights how prompt design affects LLM effectiveness and encourages future research on lightweight contextual features that are more focused. This could boost LLM performance without complicating input, resulting in more reliable software engineering applications.

% \peter{I think we need a short discussion (like 3 to 4 sentences) to discuss the implication of the finding to elevate the RQ. Checkout Large
% Language Models Can Be Easily Distracted by Irrelevant
% Contex, and cite/discuss HITS: High-coverage LLM-based Unit Test Generation via Method Slicing, and say perhaps we need to combine traditional static analysis to distill the info to better leverage LLM}

\rqbox{Including additional context reduces the accuracy of log level suggestions and increases the likelihood of invalid outputs. This suggests that concise prompts focused on relevant log messages and code are more effective, while overloading the model with undistilled context can negatively impact its performance.}

\subsection{RQ3: How is the generalizability of LLMs in suggesting log levels?}
\phead{Motivation. } %In previous sections, we have established that albeit LLMs did not exceed the standards set by SOTA, their performance was notably close. 
%LLMs excel at suggesting log levels based on learned logging patterns within the same system. Howwever, 
Newly developed software systems may lack sufficient logging statements and source code to fine-tune an LLM adequately. In such cases, fine-tuning the LLM with a limited dataset can lead to suboptimal results. Therefore, leveraging data from other projects to complement the existing dataset for fine-tuning the LLM becomes essential. Such transfer learning techniques are commonly employed to overcome the challenge of limited datasets~\cite{transferable}.
Previous studies have produced mixed findings on generalizability. DeepLV~\cite{DeepLV} reported less than a 1\% increase in all metrics from enlarging or combining datasets, while other studies~\cite{TeLL, MultiComponentLogLevel} reported different outcomes. This discrepancy motivates us to investigate whether LLMs exhibit varying degrees of transferability when subjected to different dataset and combinations.

In this RQ, we study whether transfer learning among different studied systems is suitable for LLM with the following two sub-RQs:

\noindent \textbf{RQ3-A: } Can the performance improve by including training data from other studied systems?

\noindent \textbf{RQ3-B: } How accurate are LLMs in cross-system suggestions?

%The methodologies for each sub-RQ are outlined below.

\noindent \subsubsection{RQ3-A: Can the performance improve by including training data from other studied systems}
\mbox{} % 

\phead{Approach. } We expand the training dataset by consolidating all training data from the nine Java systems. Employing stratified sampling for each system, we partition the data into training (60\%), validation (20\%), and testing (20\%) sets. Subsequently, we consolidate the training data from all systems and utilize it for fine-tuning the LLM. Meanwhile, the combined 20\% validation dataset is employed to validate the model throughout the training phase. Finally, we utilized the fine-tuned LLM trained with the expanded dataset to the testing data of each system under study.

\subsubsection{RQ3-B: How accurate are LLMs in cross-system suggestions}
\mbox{} % 

\phead{Approach. }For each target system, the training data from the remaining eight systems is consolidated and subjected to stratified sampling~\cite{stratified} with the 60\%:20\%:20\% ratio. Subsequently, this training data is utilized to fine-tune the LLM and is tested on the remaining test data from the other eight systems.

\begin{table*}
\caption{Comparison between Cross Dataset Training and Enlarged Dataset Training Fill-Mask (GraphCodeBERT) and Text Generation (CodeLlama 7B) LLMs in terms of Accuracy, AUC and AOD.}
\centering
\resizebox{2.05\columnwidth}{!}{%
\begin{tabular}{lcccccccccccc}
\toprule
\multirow{3}{*}{Project} & \multicolumn{6}{c}{Cross Dataset Training} & \multicolumn{6}{c}{Enlarged Dataset Training} \\
\cmidrule(lr){2-7} \cmidrule(lr){8-13}
& \multicolumn{3}{c}{Fill-Mask} & \multicolumn{3}{c}{Text Generation} & \multicolumn{3}{c}{Fill-Mask} & \multicolumn{3}{c}{Text Generation} \\
\cmidrule(lr){2-4} \cmidrule(lr){5-7} \cmidrule(lr){8-10} \cmidrule(lr){11-13} 
& Acc. & AUC & AOD & Acc. & AUC & AOD & Acc. & AUC & AOD & Acc. & AUC & AOD \\
\midrule             
Cassandra & 46.7 (-18.7) & 73.3 (-9.4) & 72.2 (-9.7) & 38.8 (-7.1) & 70.9 (+2.0) & 72.5 (-1.8) & 55.5 (-9.8) & 76.9 (-5.8) & 76.4 (-5.5) & 45.8 (-0.1) & 75.3 (+6.4) & 74.3 (+0.0) \\
ElasticSearch & 39.0 (-29.0) & 67.6 (-14.7) & 68.5 (-14.0) & 40.7 (+11.8) & 69.2 (+8.4) & 71.0 (+8.6) & 42.5 (-25.5) & 71.0 (-11.3) & 72.6 (-9.9) & 47.9 (+19.0) & 74.5 (+13.7) & 76.8 (+14.4) \\
Flink & 52.8 (-19.2) & 78.5 (-9.0) & 73.9 (-11.9) & 53.8 (+3.0) & 78.0 (+3.6) & 75.3 (-3.1) & 56.2 (-15.8) & 81.3 (-6.2) & 76.9 (-8.8) & 60.3 (+9.5) & 81.2 (+6.8) & 78.1 (-0.3) \\
Hbase & 51.0 (-21.0) & 77.6 (-8.8) & 74.5 (-11.0) & 48.7 (-3.4) & 75.7 (+2.0) & 74.4 (-3.4) & 54.0 (-18.0) & 78.6 (-7.8) & 76.7 (-8.8) & 55.6 (+3.5) & 80.2 (+6.5) & 79.0 (+1.2) \\
Jmeter & 54.4 (-23.6) & 79.1 (-10.4) & 77.0 (-12.4) & 53.7 (+7.3) & 78.2 (+7.7) & 77.5 (+1.8) & 60.4 (-17.6) & 81.2 (-8.3) & 81.2 (-8.2) & 61.3 (+14.9) & 81.0 (+10.5) & 80.8 (+5.1) \\
Kafka & 47.7 (-18.1) & 75.3 (-8.1) & 72.6 (-10.5) & 44.3 (+6.6) & 74.1 (+8.7) & 73.6 (+5.7) & 46.9 (-18.9) & 74.9 (-8.5) & 71.7 (-11.4) & 47.8 (+10.1) & 77.7 (+12.3) & 75.6 (+7.7) \\
Karaf & 59.0 (-5.1) & 83.3 (+0.1) & 80.3 (-1.1) & 46.4 (-0.3) & 78.9 (+4.0) & 79.0 (+0.7) & 61.8 (-2.3) & 83.7 (+0.5) & 81.8 (+0.4) & 51.9 (+5.2) & 81.9 (+7.0) & 80.7 (+2.4) \\
Wicket & 48.4 (-21.1) & 74.0 (-6.6) & 71.6 (-11.0) & 35.3 (-5.4) & 69.8 (+2.7) & 68.1 (-3.5) & 52.4 (-17.1) & 77.2 (-3.4) & 74.9 (-7.7) & 37.9 (-2.8) & 70.8 (+3.7) & 66.5 (-5.1) \\
Zookeeper & 47.9 (-14.5) & 79.6 (-3.1) & 74.7 (-7.3) & 37.0 (-12.2) & 74.6 (+1.1) & 73.9 (-2.3) & 52.0 (-10.4) & 81.3 (-1.4) & 77.0 (-5.0) & 42.1 (-7.1) & 77.4 (+3.9) & 76.6 (+0.4) \\
\midrule
\textit{Average} & 49.7 (-18.9) & 76.5 (-7.8) & 73.9 (-9.9) & 44.3 (+0.0) & 74.4 (+4.5) & 73.9 (+0.3) & 53.5 (-15.1) & 78.5 (-5.8) & 76.6 (-7.2) & 50.1 (+5.8) & 77.8 (+7.9) & 76.5 (+2.9) \\
\bottomrule
\end{tabular}%
}
\label{table:RQ3}
\begin{tablenotes}
     \item \footnotesize{Note: The +/- number after each data denotes the relative improvement or decline compared to the suggestions within the system as observed in RQ1.} 
   \end{tablenotes}
\vspace{-5pt}  
\end{table*}

\phead{Results. } \textit{\textbf{LLMs do not benefit from transfer learning, whether through enlarging or combining datasets.}} Table~\ref{table:RQ3} shows the results of enlarging the training set and combining the training set. We compared the performance of the best Fill-Mask and Text Generation LLMs identified in RQ1 with their within-system suggestions noted in RQ1, highlighting the differences in the values provided in parentheses. 
Overall, both cross and enlarged dataset training led to decreases in accuracy, AUC, and AOD for both LLM types. We noticed that Fill-Mask LLM performance decreased more significantly compared to the Text Generation LLM. The accuracy dropped 15.1\% and 11.3\% by Fill Mask LLM in cross and enlarged dataset training respectively, whereas Text Generation LLMs' accuracy reduced 14.0\% and 8.2\%. Despite Fill Mask LLM suffering more, the average values for all three metrics still surpassed the performance of Text Generation LLM. 

\textit{\textbf{Larger datasets and fine-tuning have limited impact on Text Generation LLM Performance for Log Level Suggestion.}} 
% Another observation is that the accuracy, AUC, and AOD of the Text Generation LLM appear to be less influenced by enlarging or combining the datasets. This observation aligns with the findings in RQ1, where increasing the number of sampling shots did not significantly enhance the performance of Text Generation LLMs.
Another observation is that the accuracy, AUC, and AOD of Text Generation LLMs are less affected by enlarging or combining datasets. This aligns with RQ1 findings, where increasing the number of sampling shots did not significantly improve Text Generation LLM performance.

% Table~\ref{table:RQ3} there is an average decrease of 18.9\% in Accuracy for Fill-Mask LLMs and 4.0\% for Text Generation LLMs, proving that fine-tuning for LLMs do not benefit from the enlarged dataset. This finding aligns with previous studies~\cite{TeLL, MultiComponentLogLevel} where cross-system suggestions demonstrate lower performance as compared to within-system suggestions. 
Table~\ref{table:RQ3} shows an average decrease of 18.9\% in Accuracy for Fill-Mask LLMs and 4.0\% for Text Generation LLMs, indicating that fine-tuning LLMs does not benefit from the enlarged dataset. This finding aligns with previous studies~\cite{TeLL, MultiComponentLogLevel}, where cross-system suggestions performed worse than within-system suggestions.

\rqbox{LLMs excel in suggesting log levels within the same system rather than across different systems, emphasizing the importance of specific data contexts in log level suggestion.}
\section{Implications and Future Works}
\label{sec:futureworks}
Based on our empirical findings, we highlight actionable insights and future directions for two key audiences: \circled{1} developers, and \circled{2} researchers.

\subsection{Developers}

% \peter{we need to give a recommendation on what models to use for log level suggestion, and how our finding can be applied to other code enhancement tasks}

%\phead{Enhanced Log Level Suggestions.} In our research, we concentrate on leveraging the most readily available information to developers: the source code of methods for suggesting log levels. However, our results did not surpass the current state-of-the-art (SOTA). Future investigations could explore whether integrating these features with LLMs could establish the next benchmark in log-level suggestion. By integrating LLMs with the features derived from techniques such as DeepLV~\cite{DeepLV} and TeLL~\cite{TeLL}, we could potentially achieve a new benchmark that improves the accuracy of log level suggestions, improving debugging and logging practices.
 
\phead{Approaches for Utilizing LLMs to Enhance Code Improvement Tasks.}
We identified key strategies for using LLMs to improve log level suggestions, which can also benefit other code enhancement tasks. While larger LLMs often provide better accuracy, smaller, code-specific models can achieve similar results with less resource use, making them more efficient. When resources are limited, Fill-Mask LLMs can also perform well despite their size. It's important to choose an LLM trained on relevant data for specific logging tasks, as this improves performance more than simply selecting larger models. Developers should focus on fine-tuning to enhance accuracy instead of relying on in-context learning. Although Text Generation LLMs show strong initial performance, they can be hard to fine-tune, so exploring other models or techniques may lead to better results. 
% These strategies can improve log level suggestions and support other code enhancements.
By following these strategies, developers can improve log level suggestions and support other code enhancement efforts.

\subsection{Researchers}

\phead{Enhancing LLMs through Information Slicing and Additional Code Features.}
In Section~\ref{sec:results}, we found that short prompts decrease the risk of invalid outputs in log level suggestions. Li et al.~\cite{zhenhaoICSE2023} emphasize the importance of dynamic variables in logging, noting that log variables are typically shorter than calling method code. Additionally, Wang et al.~\cite{wang2024hitshighcoveragellmbasedunit} discovered that breaking information into slices improves LLM performance in generating test cases. By combining these findings, we can explore the potential of using shorter log variables alongside information slicing to enhance LLM effectiveness. Future research could utilize static code analysis to identify and incorporate additional code features, such as log variables, which may further improve performance.

\phead{Effective LLMs Are Not Just About Parameters.}
In our study, we assessed 12 open-source LLMs for log level suggestion and identified several performance patterns that provide insights for selecting the most effective models and optimizing their use. We found that larger parameters do not always perform better; instead, smaller, specialized models can be equally or more effective. Based on these insights, future research should focus on utilizing diverse training tasks, integrating dynamic variables, and employing information slicing techniques to further enhance LLM performance. Additionally, leveraging static code analysis to identify relevant code features may further improve model effectiveness.

% \peter{this is good, but kind of duplicate with the one below. Please try to merge the two}
\phead{Expanding the Scope of Training Tasks and Increasing Feature Diversity.}
To enhance the overall performance of LLMs in software engineering, it is essential to evaluate the incorporation of a diverse range of software development tasks beyond log level suggestion, including code enhancement, refactoring, and testing. Integrating these varied challenges into the training process allows LLMs to acquire a more profound understanding of coding practices and their complexities. This expanded training set will strengthen the models' robustness and versatility, enabling them to better meet the diverse needs of developers. Future research should prioritize this inclusion, ultimately leading to the development of more effective LLMs that can better support developers and software teams in real-world applications.
\section{Threats to Validity}
\label{sec:threats}

\phead{Construct Validity.}
 Our methodology assumes that the training data consists of high-quality source code that follows best logging practices. However, there is no universally accepted industrial standards for writing logging statements. To mitigate the issue, %For this study, 
 we selected large-scale, well-maintained systems of varying sizes across different domains that have been widely %used in previous research on logging practices and are 
 recognized by prior studies for their adherence to established logging standards~\cite{DeepLV, TeLL, hengli2018}. We evaluate our models using the test datasets from each of these systems. It is important to note that different test datasets can yield diverse outcomes. To mitigate the impact of this variability, we employ stratified random sampling techniques, as utilized in prior studies~\cite{DeepLV, DLFinder, 8840982, stratified}, to partition the dataset while maintaining the original dataset's distribution of labels.

\phead{Internal Validity.}
Randomness are observed during the inference process of LLMs. To mitigate this threat, we regulate the model temperature to 0, ensuring that LLMs consistently yield more consistent outputs for identical input text. 

\phead{External Validity.}
The nine subjects of this study are open-source Java projects from the Apache Software Foundation. While the coding style specific to Apache may limit the applicability of our findings to other organizations, the study spans various domains, project sizes, and logging volumes for broader representativeness. Although evolving software practices and languages like C++ could affect the findings, our 0-shot experiments show minimal data leakage concerns, as low accuracy suggests the base model’s unfamiliarity with the specific systems. 
\section{Conclusion}
\label{sec:conclusions}

In this study, we examined log level suggestion across nine Java systems using twelve open-source LLMs, focusing on leveraging readily available data like method source code and log messages. We found that LLM performance varies significantly by task and model type, with Code-based LLMs generally outperforming NLP-based LLMs for log level suggestion. Text Generation LLMs excelled with few-shot prompting, while Fill-Mask LLMs responded better to fine-tuning. Including the source code of calling methods decreased performance and increased invalid outputs. Our research highlights the importance of task-specific data and suggests that Text Generation LLMs are preferable when such data is unavailable. These findings offer valuable insights for enhancing LLMs in log level suggestion and guiding future research on refining LLMs for various code-related tasks.

\balance
% ------------reference------------
% \bibliographystyle{IEEEtran}
% \bibliographystyle{ACM-Reference-Format}
\bibliographystyle{IEEEtran}
\bibliography{paper}
% \bibliography{output.bbl}

\end{document}